\documentclass{ws-procs975x65}
\usepackage{psfrag}
\newcommand{\be}{\begin{equation}}
\newcommand{\ee}{\end{equation}}
\newcommand{\bea}{\begin{eqnarray}}
\newcommand{\eea}{\end{eqnarray}}
\newcommand{\T}{{\rm T}}

\newcommand{\W}{W}
\newcommand{\kap}{{\kappa}}

\newcommand{\R}{{\rm R}}

\begin{document}



\title{GRAVITATING NON-ABELIAN SOLITONS AND HAIRY
BLACK HOLES IN HIGHER DIMENSIONS}

\author{MIKHAIL S. VOLKOV}

\address{Laboratoire de Math\'ematiques et Physique Th\'eorique,
Universit\'e Fran\c{c}ois-Rabelais Tours,
F\'ed\'eration Denis Poisson - CNRS,
Parc de Grandmont, 37200 Tours, FRANCE\\
\email{volkov@lmpt.univ-tours.fr}
}


\begin{abstract}
This is a short review of classical 
solutions with gravitating Yang-Mills fields
in $D>4$ spacetime dimensions. The simplest 
SO(4) symmetric particlelike and SO(3) symmetric vortex type solutions 
in the  Einstein-Yang-Mills theory in $D=5$ are considered, and their
various generalizations with or without an event horizon, 
for other symmetries, in more general theories, and also in $D>5$ are described.   
In addition, supersymmetric solutions with gravitating Yang-Mills fields
in string theory are discussed. 
 
\end{abstract}

\bodymatter

\section{Introduction}\label{intro}
In this mini-review the following two questions will be addressed:
\begin{itemize}

\item 
What is known about solutions with gravity-coupled non-Abelian gauge fields in $D>4$ 
spacetime dimensions ?

\item 
Why should one study such solutions ?

\end{itemize}
In answering these questions we shall briefly consider 
the simplest  gravitating non-Abelian solutions   in $D=5$
and shall then discuss their  known generalizations.  

\section{Einstein-Yang-Mills theory in D=4}

Before going to higher dimensions, it is worth reminding the situation in 
$D=4$. It is well known that neither pure gravity nor pure 
Yang-Mills theory in $D=4$
admit solitons, by which we mean stationary and globally regular 
(but not necessarily stable) 
solutions with finite total energy. 

In the case of pure Einstein gravity
with the action  
\be
S_{\rm E}=\frac{1}{16\pi G}\int R\,\sqrt{-g}\,d^4 x
\ee
the statement is the content of the theorem \cite{Lich} of 
Lichenrowitcz. The vacuum gravity admits however 
solutions describing localized
stationary objects with finite mass, but these are not globally regular: 
black holes.

The action for the pure Yang-Mills (YM) theory with a compact and simple gauge group
${\cal G}$
is given by 
\be
S_{\rm YM}=-\frac{1}{4Ke^2}\,\int {\rm tr}\, F_{\mu\nu}F^{\mu\nu}\sqrt{-g}\,d^4 x\,,
\ee
where $F_{\mu\nu}=\partial_\mu A_\nu-\partial_\nu A_\mu
-i[A_\mu,A_\nu]$. Here $A_\mu=A^a_\mu \T_a$ and $\T_a$ are the gauge group
generators normalized such that ${\rm tr}(\T_a\T_b)=K\delta_{ab}$. If
${\cal G}$=SU(2)
then $a=1,2,3$ and $\T_a=\frac12\,\tau_a$ where $\tau_a$ are the Pauli matrices. 
The gauge coupling constant is denoted by $e$. 
Since this theory is scale invariant, the non-existence of stationary solitons,
expressed by the statement ``there are no classical glueballs'',   
essentially follows  \cite{Deser,Coleman,Coleman1} from the scaling arguments. 

The physical reason for the non-existence of solitons in the above two cases 
is clear: one cannot have equilibrium objects 
in systems with only attractive (gravity) or only repulsive (Yang-Mills)
interactions.  However, in systems with both gravity and Yang-Mills
fields described by the Einstein-Yang-Mills (EYM) action 
\be             \label{EYM}
S_{\rm EYM}=S_{\rm E}+S_{\rm YM}
\ee
both attractive and repulsive forces are present, and so the 
existence of solitons is not excluded. At the same time, the 
existence of solitons in this case is not guaranteed either,
and so it was a big surprise when 
such solutions were constructed \cite{BK}
by Bartnik and McKinnon. As this was actually 
the first known example of solitons in self-gravitating systems,
a lot of interest towards the 
Einstein-Yang-Mills system \eqref{EYM} has been triggered. This interest was  
further increased by the discovery in the same system of the first known example 
\cite{MV89,Kunzle,Bizon}
of hairy black holes  which are not uniquely characterized by their 
conserved charges and so violate manifestly the no-hair conjecture \cite{Wheeler}.
Further surprising discoveries followed when 
these static EYM solitons and black holes were generalized \cite{Kunz,Kunz1,Kunz2}
to the non-spherically symmetric and 
non-static cases. 
This has revealed that the EYM solutions provide manifest counter-examples also 
to a number of classical theorems  originally proven 
for the Einstein-Maxwell system but believed to be generally true, as for example 
the theorems of staticity, circularity, 
the uniqueness theorem, Israel's theorem  an so on (see \cite{MV98} for a review). 
As a result, the EYM theory has become an interesting topic of studies.

\section{Pure gravity and pure Yang-Mills in D=5}

Going to higher dimensions, let us first consider separately 
some simplest solutions for pure gravity and pure
Yang-Mills theory in $D=5$. 
We shall denote the spacetime coordinates by $x^M$ where 
$M=0,1,2,3,4$.

 \subsection{Pure gravity}
Vacuum gravity in $D>4$ has been much studied and many interesting solutions have 
been obtained (see \cite{Reall} for a recent review), but we shall just mention a couple
of the simplest ones.   
If $g^E_{\mu\nu}(x^\sigma)$, where $\mu,\nu,\sigma=1,2,3,4$, is a 
{Euclidean} Ricci flat metric 
(gravitational instanton), then the 5D geometry 
$$
ds^2=-dt^2+g^E_{\mu\nu}dx^\mu dx^\nu
$$
will also be Ricci flat. One obtains in this way 
static and globally regular vacuum solutions in $D=5$. However, 
since there are no \cite{eguchi} asymptotically Euclidean gravitational instantons, 
these 5D solutions will not be asymptotically flat (although they can be 
asymptotically locally flat). 
There exist also 
localized and asymptotically flat solutions -- black holes -- but these are
not globally regular. 
In the simplest case with the SO(4) spatial symmetry this is the Schwarzschild
black hole,
 $$
ds^2=-Ndt^2+\frac{dr^2}{N}+r^2d\Omega_3^2,~~~~
N=1-\left(\frac{r_g}{r}\right)^2. 
$$
If  $\partial/\partial x^4$ is a symmetry 
and $g_{\mu\nu}$ is a  Ricci flat Lorentzian metric /$\mu=0,1,2,3$/ then  
the 5D metric 
$$
ds^2=g_{\mu\nu}dx^\mu dx^\nu+(dx^4)^2
$$
will also be Ricci flat. Choosing $g_{\mu\nu}$ to be a 4D black hole metric
gives then a one-dimensional linear mass distribution in $D=5$ -- black string.

\subsection{Pure Yang-Mills theory: particles and vortices}
Since the Yang-Mills theory in $D=5$ is not scale invariant, 
one can have  localized and globally  regular object made of pure gauge field.
We shall call them `Yang-Mills particles', these are 
essentially the 4D YM instantons uplifted to D=5.  
 Specifically, 
if $A^a_\mu(x^\nu)$ is a solution of the 4D Euclidean 
YM equations then 
$$
A^a_M=(0,A^a_\mu(x^\nu))~~~~~~~~/\mu=1,2,3,4/
$$
will be a solution of the 5D YM equations describing a static object 
whose 5D energy coincides with the 4D instanton action \cite{inst}
$$
E=\frac{1}{4e^2}\int (F^a_{\mu\nu})^2 d^4x\geq \frac{8\pi^2 |n|}{e^2}.
$$
The equality here is attained for self-dual configurations, with 
$F_{\mu\nu}=\ast F_{\mu\nu}$, 
in which case 
$n\in\mathbb{Z}$
is the number of the `YM particles'. 
Their mutual interaction forces being exactly zero, 
the particles can be located anywhere in the 4-space.   

If 
$\partial/\partial x^4$ is a symmetry, then choosing 
$$
A^a_M=(0,A^a_i(x^k),H^a(x^k)),~~~~~~/i,k=1,2,3/
$$
the YM energy per unit $x^4$,
$$E=\frac{1}{2e^2}\int(
(\partial_iH^a+\varepsilon_{abc}A^b_i H^c)^2+\frac12 (F^a_{ik})^2)d^3 x
\geq \frac{4\pi|n|}{e^2}\,,
$$
coincides with 
the energy \cite{mon} of the D=3 YM-Higgs system. Its absolute minima for a given $n$
are the BPS monopoles, and 
when lifted back to D=5 they become one-dimensional objects --
{`YM vortices'}. The number of vortices, $n$, can be arbitrary, and they 
can be located anywhere in the 3-space of $x^k$. 

We shall now describe the self-gravitating analogs \cite{MV01}  
for the YM particles and also for the YM vortices. Practically all other known solutions with 
gravitating gauge fields in higher dimensions are generalizations of these two simplest 
types.

\section{Gravitating YM particles \cite{MV01} }
The EYM theory with gauge group SU(2) in D=5 is defined by the action 
\be                         \label{0}
S=\int{\cal L}_5\sqrt{^{(5)} g}\,d^5 x=
\int\left(\frac{1}{16\pi G}\,^{(5)}R
-\frac{1}{4 e^2}\,F^a_{MN}F^{aMN}\right)\sqrt{^{(5)}g}\,d^5 x\, ,
\ee
where 
$F^a_{MN}=\partial_M A^a_N-\partial_N A^a_M
+\varepsilon_{abc}A^b_M A^c_N$ $(a=1,2,3)$. The Newton constant $G$ 
and gauge coupling $e$ being both dimensionful,
$[G^{1/3}]=[{ e^2}]=[{\rm length}]$, one can define the 
dimensionless coupling parameter as
$$
{\kappa}=\frac{8\pi G}{ e^6}. 
$$
Let us try to construct the 5D counterparts of the 4D solitons \cite{BK}
of Bartnik and McKinnon: particle-like objects localized in 4 spatial
dimensions. In the simplest case they are static and purely magnetic
 and have the SO(4)
spatial symmetry. The fields are then given by 
\be        \label{e0}
ds^2=e^2\left\{-\sigma(r)^2 N(r)dt^2
+\frac{d r^2}{N(r)}+
r^2\,d\Omega_3^2\right\}\, ,~~~A^a=(1+w(r))\,\theta^a\, ,
\ee
where $\theta^a$ are the 
invariant forms on $S^3$. 
With $N\equiv 1-{ \kappa} m(r)/r^2$,
the  EYM field equations for the action \eqref{0} reduce to
\bea          
&&r^2 Nw''+r\,w'
+{ \kappa}\,(m-(w^2-1)^2)\frac{w'}{r}=2\,(w^2-1)w\, , \label{e1} \\
&&rm'=r^2Nw'^2+(w^2-1)^2\, ,~~   \label{e2} \\
&&\sigma'=\kap\,\frac{w'^2}{r}\,\sigma\,~~ \Rightarrow~~
\sigma(r)=\exp\left(-{ \kappa}\int_r^\infty\frac{w'^2}{r}\,dr\right).
\label{e3}
\eea
If ${\kappa=0}$ then $\sigma=N=1$ and the geometry is flat. 
 The solution of 
Eq.\eqref{e1} is 
the 5D { YM particle}, which is the same as the 4D one-instanton solution:
\be                                 
w=\frac{1-{ b}\,r^2}{1+{ b}\,r^2}.
\ee
Here ${ b}>0$ is an integration constant that determines the 
size of the particle. 
This solution is everywhere regular and its 
energy is the ADM mass: $E=M_{\rm ADM}=m(\infty)=\frac{8}{3}$. 

The next question is what happens if ${\kappa\neq 0}$ ?
Let us consider asymptotically flat solutions with finite ADM mass.
If they are globally regular, then they will have a regular origin at $r=0$,
and for small $r$ the local solution of Eqs.\eqref{e1}--\eqref{e3} will be given by 
\be                     \label{reg} 
w=1-2br^2+O(r^4),~~~~~m=O(r^3).
\ee
One can also consider black hole solutions, in which case  
\be                                   \label{bh} 
\exists r_h>0:~~~~~N(r_h)=0,~~~~~N'(r_h)>0,~~~~~w(r_h)<\infty\,.
\ee
In both cases, using the boundary condition \eqref{reg} if $r_h=0$ and  
\eqref{bh} for $r_h>0$ and 
integrating \eqref{e2} gives the ADM mass functional 
\be                                            \label{mass}
M_{\rm ADM}[w(r)]=m(\infty)
=\frac{r_h^2}{\kappa}+
\int_{r_h}^\infty\frac{dr}{r}\,(r^2w'^2+(w^2-1)^2)
\sigma(r). 
\ee
This shows that one should have $w(\infty)=\pm 1$ for the 
mass to be finite. Let us suppose that there is a solutions with $M_{\rm ADM}<\infty$
and let $w(r)$ be the corresponding gauge field amplitude.  
The first variation 
of the mass functional \eqref{mass} should then vanish under all  variations 
$w(r)\to w(r)+\delta w(r)$ 
respecting the boundary conditions.  
In particular, if we consider special variations $w(r)\to w(\lambda r)$ then 
one should have 
$\frac{d}{d\lambda}M[w(\lambda r)]=0$ for $\lambda=1$. 
However, \eqref{mass} implies that 
$\frac{d}{d\lambda}M[w(\lambda r)]<0$ for any $\lambda$. As a result, 
there are no solutions with finite ADM mass.

We therefore conclude that
{\sl the flat space YM particles do not admit self-gravitating generalizations
with finite energy}. One can say that they  
{ get completely destroyed by gravity}, as in some sense they  
resemble dust, since their energy momentum tensor is 
$$
T_{MN}=\epsilon(r)\delta^0_M\delta^0_N
$$
and in addition 
they can be scaled to an arbitrary size. As a result, when gravity is on, 
the repulsion and attraction are not balanced and so  equilibrium states
are not possible. 
If one nevertheless tries 
integrating Eqs.\eqref{e1}--\eqref{e3} starting from the 
regular origin, say, to see what happens, one finds a 
peculiar quasi-periodic behavior shown in Fig.1. 
\begin{figure}[h]
\hbox to\linewidth{\hss%
  \resizebox{8cm}{6cm}{\includegraphics{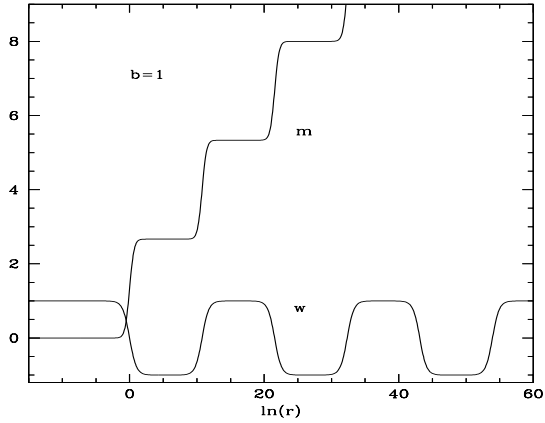}}%
\hss}
\caption{Globally regular solution to Eqs.\eqref{e1}--\eqref{e3},\eqref{reg}
with $b=1$, $\kappa=10^{-8}$. }
\end{figure}
This can be explained qualitatively:
integrating  Eq.\eqref{e1} with the regular boundary condition \eqref{reg}
gives
\be               \label{11}
w^{\prime 2}-(w^2-1)^2=
-\int_{-\infty}^\tau(\ln(\sigma^2 N))'w'^2 d\tau\, ,
\ee
with $^\prime \equiv\frac{d}{d\tau}\equiv r\sqrt{N}\frac{d}{dr}$.
This describes a particle moving with friction
in the inverted double-well potential $U=-(w^2-1)^2$. 
At $\tau=-\infty$ the particle
starts at the local maximum of the potential
at $w=1$, but then it looses energy 
due to the dissipation and
cannot reach the second maximum at $w=-1$. 
Hence it bounces back.
For large $\tau$ 
the dissipative term tends to zero, 
and the particle ends up oscillating in the
potential well with a constant energy. After each oscillation the mass
function $m$ increases in a step-like fashion, and for large $r$
one has $m(r)\sim \tau\sim\ln r$. As a result, 
there emerges an infinite sequence of static spherical shells of the 
YM energy in the D=5 spacetime. These arguments easily generalize to the black hole case.

\subsection{Other particlelike solutions \cite{Maeda,
BCT-High,BCHT-High,RT-High,
RST-High,BRT-High,BMT,RST
}}

A number of generalizations 
of the above results have been considered. It
turns out that the divergence of the mass is the generic property 
of the static EYM solutions with the spherical symmetry in $D>4$. 
In particular, it has been shown that for all static 
EYM solutions with gauge group ${\cal G}=$SO(d) and with the spatial SO(d) symmetry
in $D>4$ dimensions, 
where $d=D$ if $D$ is even and $d=D-1$ if $D$ is odd, the ADM mass diverges. 
This concerns both globally regular and black hole solutions, and also 
solutions with a $\Lambda$-term. 

It turns out, however, that one can nevertheless have finite mass particlelike 
solutions in $D>4$ at the expense of adding 
the higher order curvature corrections to  the EYM Lagrangian.  
The gravitational term of the action, $R$, is then `corrected' by adding all possible
Gauss-Bonnet invariants, while the Yang-Mills term, ${\rm tr}(F\wedge \ast F)$,
is augmented by adding terms of the type  ${\rm tr}({\cal F}\wedge \ast {\cal F})$,
where ${\cal F}=F\wedge\ldots\wedge F$.  
It is important that after taking these corrections into account the field equations remain 
the second order. These equations admit localized solutions with a finite mass which have 
been studied \cite{
BCT-High,BCHT-High,RT-High,
RST-High,BRT-High,BMT,RST
}
quite extensively.

\subsection{Gravitating Yang monopoles \cite{GT}} 
Flat space 4D YM instantons with ${\cal G}$=SU(2) 
can also be used to construct self-gravitating solutions 
in $D=6$. The YM instanton lives in this case on the $S^4$ angular part of 
the SO(5)-invariant 5-space, 
the ansatz for the metric and gauge field being 
$$
ds^2=-\sigma(r)^2 N(r)dt^2
+\frac{d r^2}{N(r)}+
r^2(d\xi^2+\sin^2\chi\,d\Omega_3^2)\, ,~~~A^a=(1+w(\chi))\,\theta^a\, . 
$$
YM equations decouple and admit a solution $w(\chi)=\cos\chi$, 
whose energy momentum tensor does not depend on $r$ and 
determines the source for the Einstein equations. 
The latter give  
$$
\sigma=1,~~~~N=1-\frac{2Gm(r)}{r^3},~~~~~m'={ 8\pi},~~~~~m(r)=8\pi r+m_0,
$$
and so the mass is linearly divergent for large $r$. 
This solution can be generalized \cite{GT} also to 
$D=2k+2$ dimensions, choosing SO($2k$) as  the gauge group. 
In this case one finds $m(r)\sim r^{2k-3}$.

\section{Gravitating YM vortices \cite{MV01}}
Let us return to the 5D EYM model defined by Eq.\eqref{0} and assume the 
existence of a  
 hypersurface orthogonal Killing vector $\partial/\partial x^4$,
such that the metric is 
$$
g_{MN}dx^M dx^N=e^{-\zeta}g_{\mu\nu}dx^\mu dx^\nu+e^{2\zeta}(dx^4)^2\,,
$$
\be                                  \label{hyper} 
A^a_M dx^M=A^a_\mu dx^\mu+H^a dx^4\, ,~~~/\mu,\nu=0,1,2,3/.
\ee
Inserting this to \eqref{0} reduces the 5D EYM model to 
the 4D EYM-Higgs-dilaton theory, 
\bea                                   \label{hyper1} 
\sqrt{^{(5)}g}\,{\cal L}_5&=&
\left(\frac{^{(4)}R}{2{\kappa}{ e^6}}\,-
\frac{3}{{\kappa}{ e^6}}(\partial_\mu\zeta)^2\right.\nonumber \\
&-&\left.\frac{1}{2 e^2}e^{-2\zeta}(D_\mu H^a)^2
-\frac{1}{4 e^2}e^\zeta (F^a_{\mu\nu})^2\right)\sqrt{-^{(4)}g},\nonumber
\eea
Let us assume the {SO(3) symmetry} for the 4D fields, 
$$
e^{-\zeta(r)}ds^2=-{\rm e}^{2\nu(r)}dt^2
+dr^2+
\R^2(r)\,d\Omega_2^2,
$$
\be                    \label{vort}
A^a_k dx^k=(w(r)-1)\epsilon_{aik}n^idn^k,~~~~
H^a=n^a e^{\zeta(r)}h(r),
\ee
where $n^k$ is the unit normal to the 2-sphere, $d\Omega_2^2=dn^kdn^k$. 
The independent EYM equations can be rewritten \cite{MV01} in the form of a seven-dimensional 
dynamical system 
\be                         \label{dynam}
\frac{d}{dr}\,y_k=F_k(y_s,{\kappa})
\ee
with $y_k=\{w,w',h,h',Z=\zeta',\R,\R'\}$. Solutions of these equations 
show interesting features which can be qualitatively understood 
by studying the fixed points of the system.  
The system has the following fixed points. 

{I.  The origin:} $(w,h,Z,\R)=(1,0,0,0)$. Assuming that this fixed point is 
attained for $r=0$, the local behavior of the solution for small $r$ 
is 
\bea            \label{zero}
w=1-br^2+O(r^2),~~~ h=ar+O(r^3),~~~
Z=O(r^2),~~~ \R=r+O(r^3).   
\eea 

II. { Infinity:} $(w,h,Z,1/\R)=(0,1,0,0)$. This fixed point can be reached  
for $r\to \infty$, in which case
\bea                        \label{inf}
&&w=A\,r^{C}{\rm e}^{-r}+o({\rm e}^{-r}),
\ Z={\kappa}{Q}{r^{-2}}+O(r^{-3}\ln r),\ \nonumber  \\
&&h=1-{C}{r^{-1}}+O(r^{-2}\ln r),  \\
&&\R=r-m\ln r+m^2r^{-1}\ln r-r_0+{\gamma}{r^{-1}}
+O(r^{-2}\ln r)\, .  \nonumber
\eea
In these expressions 
$a$, $b$, $A$, $v$, $C$, $Q$, $r_\infty$, $\gamma$ 
are eight free parameters. 
The ADM mass is 
$
M_{\rm ADM}=3(C+(2+{\kappa} )Q). 
$

III. { ``Warped'' $AdS_3\times S^2$.} The fixed point values of the 
functions are expressed in this case  in terms of roots of the cubic
equation 
{$4q^3+7q^2+11q=1$} as 
\be                   \label{III}
w^2=q,~~
\R^2=\kap\frac{(11q-1)(1-q)}{(4q^2-13q+1)},~~
h^2=\frac{1-q}{R^2},~~Z^2=-\frac{4q^2-13q+1}{(4q+1)R^2}. 
\ee
Evaluating, 
\be                         \label{333}
w=0.29,~ h=\frac{1.27}{\sqrt{\kap}},~ 
Z=\pm\frac{0.31}{\sqrt{\kap}} ,~ 
\R=0.75\sqrt{\kap}\, ,
\ee
which gives an exact non-Abelian solution with
the geometry 
\be                                       \label{warped} 
ds^2={\rm e}^{2(1+\kappa h^2) Zr}dt^2-d r^2-{\rm e}^{2Zr}\,(dx^4)^2
-\R^2\,d\Omega_2^2.
\ee
Linearizing Eqs.\eqref{dynam} around this fixed point gives  
the characteristic eigenvalues
\be\left( -\frac{2.77}{\sqrt{\kap}},-\frac{2.47}{\sqrt{\kap}},
-\frac{2.12}{\sqrt{\kap}},
{ -\frac{0.61}{\sqrt{\kap}}\pm i\frac{1.24}{\sqrt{\kap}}} ,
+\frac{0.88}{\sqrt{\kap}},
+\frac{1.54}{\sqrt{\kap}}\right) .                   \label{eig} 
\ee

\begin{figure}[ht]
\hbox to\linewidth{\hss%
\resizebox{9cm}{4.5cm}{\includegraphics{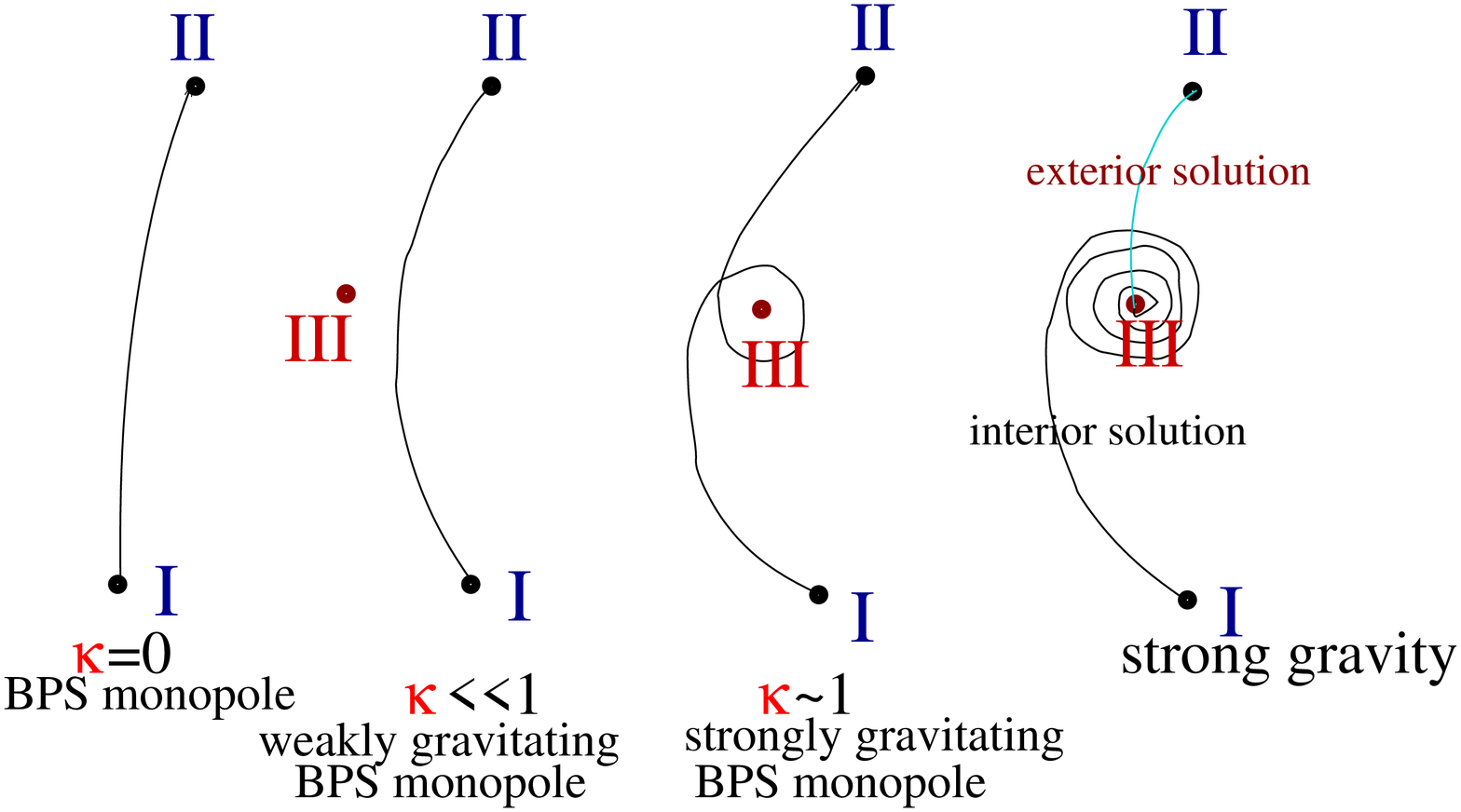}}%
\hss}
\caption{Schematic behavior of solutions of Eqs.\ref{dynam} interpolating
between fixed points I and II.}
\end{figure}  

{Global solutions} of Eqs.\eqref{dynam} can then be viewed as 
trajectories in the phase space interpolating between 
fixed point I (origin) and fixed point II (infinity). Their behavior is schematically
sketched in Fig.2. For $\kap=0$ the 4D solution is the flat space BPS 
monopole and  its 5D analog is the flat space YM vortex. For 
$\kap\ll 1$ the solution is a weakly gravitating 5D vortex whose 
field configuration is only slightly deformed as compared to the $\kap=0$ case. 
The YM vortices therefore do generalize to curved space, unlike the YM particles.  

\subsection{Strong gravity limit}
Further increasing the gravitational coupling 
$\kap$ deforms the YM vortex more and more and 
the phase space trajectory approaches closer and closer 
the third fixed point.
\begin{figure}[h]
\hbox to\linewidth{\hss%
  \resizebox{6.5cm}{5.5cm}{\includegraphics{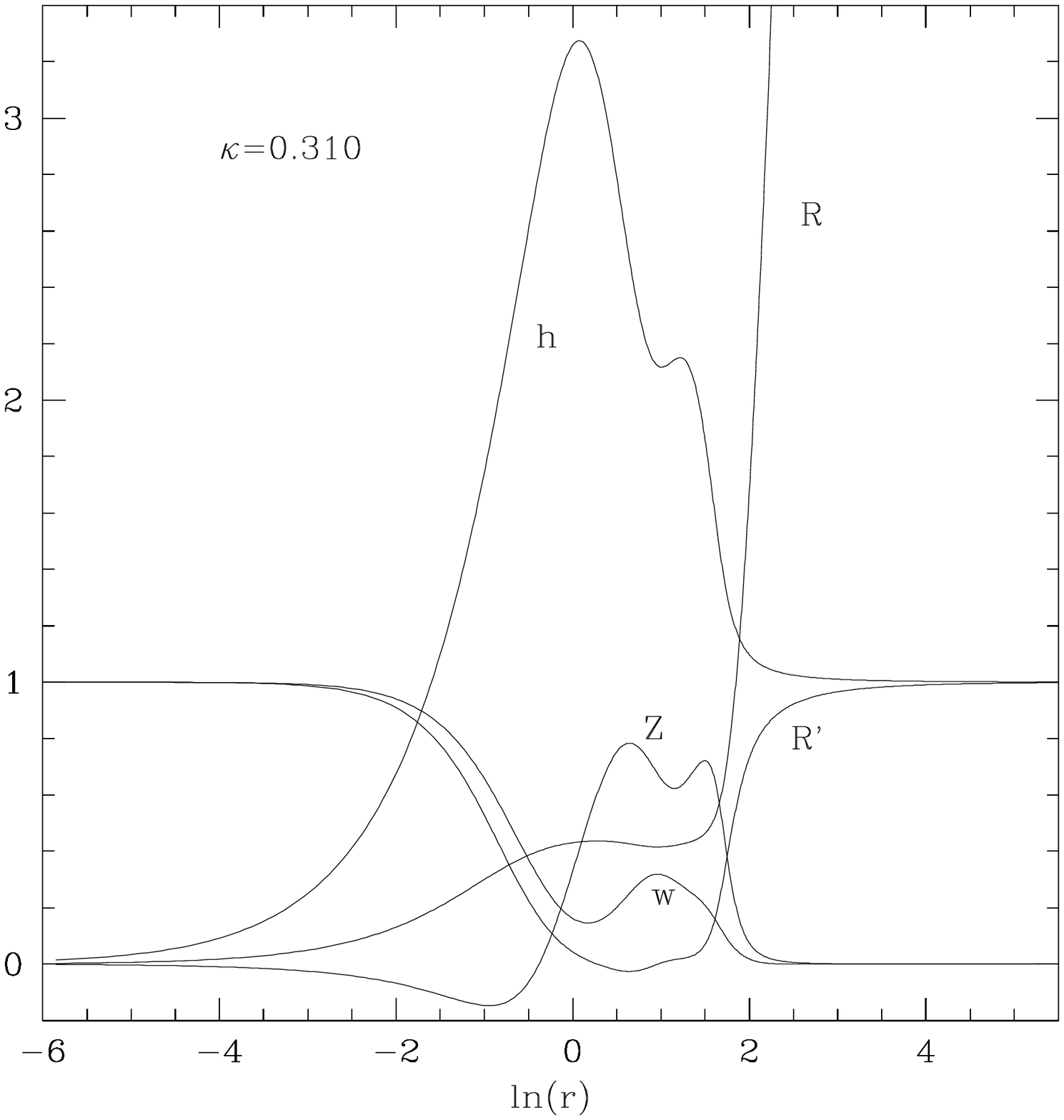}}%
  \resizebox{6.5cm}{5.5cm}{\includegraphics{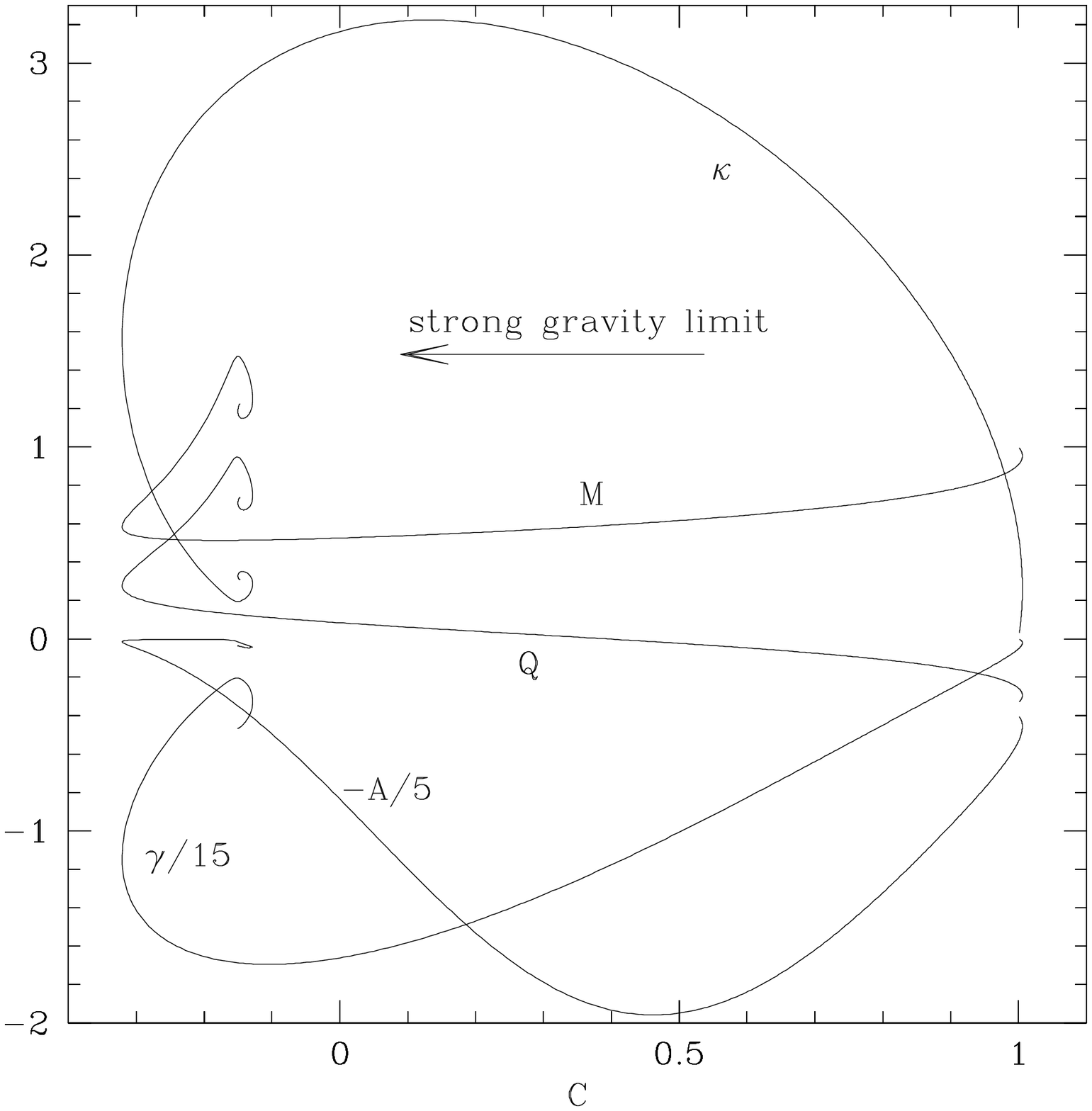}}%
\hss}
\caption{Left: For strongly gravitating solutions the field amplitudes oscillate around
the values \eqref{333} corresponding to the third fixed point. 
Right: The spiraling behavior of the solution parameters in the 
strong gravity limit. One sees that 
for a given value of $\kappa\in[0.11,3.22]$ there exist several
(at least two) different solutions.
}
\end{figure}
Since the latter has complex eigenvalues with negative real part, 
the phase trajectory gets attracted by this fixed point 
and oscillates around it for a while before going to infinity. 
These oscillations manifest themselves in the solution profiles; see Fig.3. 
A further increase of the gravitational force is not accompanied
by an increase of the value of $\kappa$ but can rather be achieved 
by increasing the value of $\xi=\kap h'(0)$. The parameters 
$\kap(\xi)$, $M_{\rm ADM}(\xi)$, etc. then 
start developing spiraling oscillations around some limiting values; see Fig.3.

Such strongly gravitating solutions have a regular core connected to 
the asymptotic region by a long throat -- the region where the radius of the 
two-sphere, $\R$, is approximately constant and all other field amplitudes 
oscillate around the constant values \eqref{333}. Increasing the value of 
$\xi=\kap h'(0)$ this throat becomes longer and longer, and finally the solution   
splits up into the two independent solutions: interior and exterior, schematically 
shown in Fig.4.  
\begin{figure}[h]
\hbox to\linewidth{\hss%
  \resizebox{6.5cm}{5.5cm}{\includegraphics{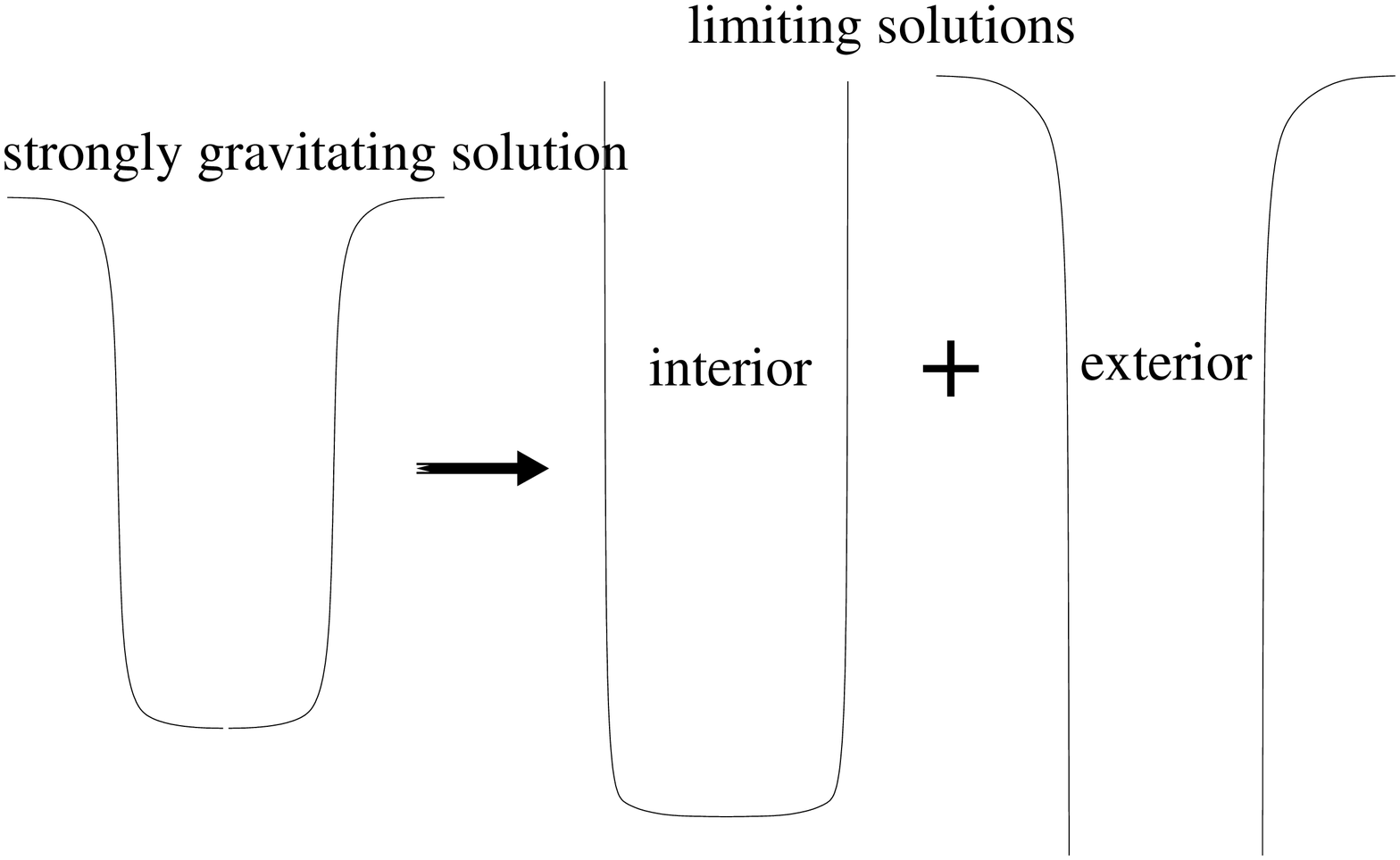}}%
  \resizebox{6.5cm}{5.5cm}{\includegraphics{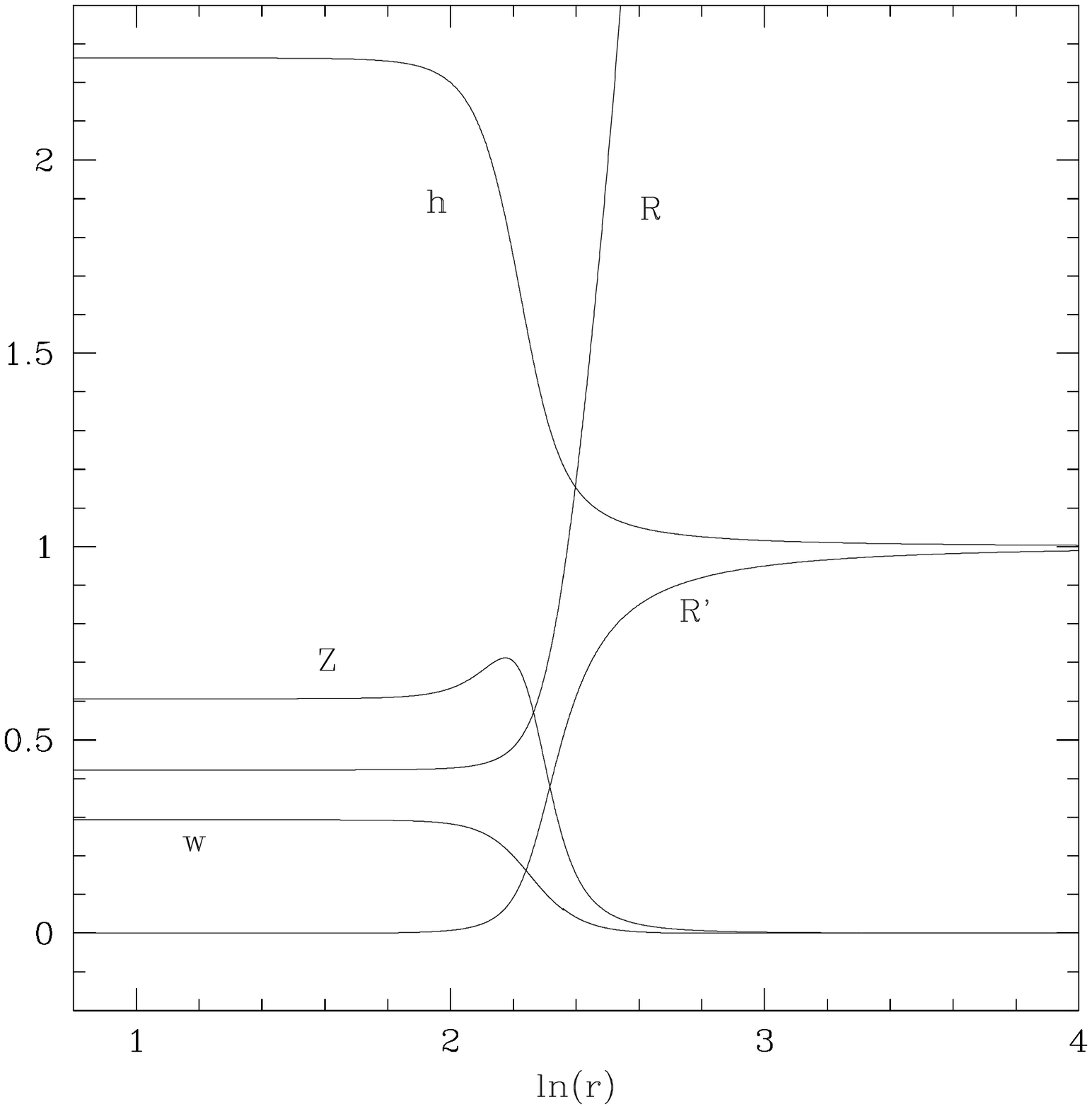}}%
\hss}
\caption{Left: Schematic behavior of solutions in the strong gravity limit.
Right: The profiles of the exterior part of the limiting solution.  
}
\end{figure}

{The interior solution} has a regular central core and approaches asymptotically 
the `warped ADS' geometry \eqref{warped}. 
In Fig.2 this solution corresponds to the phase trajectory that starts at the 
fixed point I and after infinitely many oscillations around the fixed point III 
ends up there.  {The exterior solution}, shown in Fig.4, interpolates between 
the `warped ADS' and infinity. In Fig.2 this solution corresponds to the 
phase trajectory that interpolates between the 
fixed points III and II.  It does not oscillate,
since all `repelling' eigenvalues \eqref{eig} with positive real part are real. 
This solution is somewhat similar to an extreme non-Abelian black string,
since passing to the Schwarzschild gauge where $r=\R$ one has 
$$
g_{rr}\sim (r-r_h)^{-2},~~~
g_{tt}\sim
g_{44}=e^{2\zeta}\sim (r-r_h)^{2.02}
$$
where the event horizon radius is $r_h=0.42$.

\subsection{Conclusions}
Summarizing the above discussion, although the static and SO(4) symmetric YM particles 
in D=5 get destroyed by gravity, in view of their scaling behavior,
the static and SO(3) symmetric solutions
 -- YM vortices -- do admit
non-trivial curved space generalizations, since their scale invariance is broken 
by the asymptotic value of the `Higgs field' $H^a=A^a_4$.  
These gravitating YM vortices comprise a one-parameter family of globally
regular solutions -- the fundamental branch -- that interpolates 
between the flat space BPS monopole and the extreme non-Abelian black
string. For all these solutions the YM  field amplitude $w$ is positive definite. 
In addition, there exist also excitations over the fundamental branch. These are   
solutions for which the YM field amplitude $w$ oscillates around { zero value}, these 
solutions do not have the flat space limit.

\section{Generalized YM vortices} 

The gravitating YM vortex solutions of the fundamental branch  (but not the excited ones) 
have been generalized in a number of ways.

\subsection{YM black strings \cite{Hartmann04}}
From the 4D viewpoint the YM vortices are regular 
gravitating solitons. It has been known for quite a long time \cite{Kastor92} 
that gravitating solitons can often be generalized to include a small black hole 
in the center. Technically this requires replacing 
the boundary conditions at the  regular origin (conditions of the type \eqref{reg}) 
by the boundary conditions at the regular event horizon 
 (conditions of the type \eqref{bh}). Such a procedure has been used \cite{Hartmann04} to 
 promote the regular YM vortices to  black strings. 
  
For a given $\kap$ there can be several YM vortex solutions corresponding 
to different parts of the spiraling curves in Fig.3. This property generalizes also
to the black string case, where one also finds several black string solutions 
for  a given $\kap$, provided that their even horizon radius $r_h$ is 
small enough. 
As $r_h$ increases, these
solutions approach each other and finally merge for some maximal 
value $r_h^{\rm max}(\kap)$. There are no black strings with 
$r_h>r_h^{\rm max}(\kap)$, and so black strings exist
only in a finite domain of the $(\kap,r_h)$ parameter plane.  
Black strings and regular vortex solutions with a $\Lambda$-term \cite{TB,BD}
have also been considered.

\subsection{`Twisted' solutions \cite{BR-twist,BR-twist1}}
When performing the dimensional reduction from D=5 to D=4 in Eqs.\eqref{hyper}
and \eqref{hyper1} it was assumed that the Killing vector $\partial/\partial x^4$
is hypersurface orthogonal. Relaxing this condition,  
Eq.\eqref{hyper} generalizes to
\be
g_{MN}dx^M dx^N=e^{-\zeta}g_{\mu\nu}dx^\mu dx^\nu+e^{2\zeta}
(dx^4+{ W_\mu} dx^\mu)^2\,, \nonumber 
\ee
\be
A^a_M dx^M=A^a_\mu dx^\mu+H^a (dx^4+{ W_\mu} dx^\mu)\,,\nonumber 
\ee
where the twist $W_\mu$ can be viewed as a 4D 
vector field.  
Inserting this to the action \eqref{0} gives instead of \eqref{hyper1} 
a four dimensional  EYM-Higgs-dilaton+U(1) model. The solutions of this 
`twisted' model  carry an 
additional U(1) charge under the vector field $\W_\mu$. 
This charge can be of electric \cite{BR-twist} or magnetic \cite{BR-twist1} type. 
When this 
charge vanishes, the solutions reduce to the YM vortices/black strings.

\subsection{Deformed and stationary solutions \cite{BH02a,BHR05def,BHR05}}
The 5D YM vortices/black strings, with or without twist, have been also generalized 
\cite{BH02a,BHR05def,BHR05} to the case where, after the dimensional reduction to D=4, the 
fields are chosen to be static and axially symmetric, rather than spherically symmetric.  
The ansatz for the gauge field $A_\mu^a$ and $H^a$ 
contains in this case two integer winding numbers, $n,m$. 
If ${ n=1},{ m=0}$ then the solutions are spherically symmetric monopoles. 
For ${ n>1}$, ${ m=0}$ 
one obtains axially symmetric solutions of the multimonopole type. 
Solutions with ${ m=1}$ describe monopole-antimonopole pairs. 
Both regular solutions \cite{BHR05def} and black strings \cite{BHR05} have been considered within this approach.  In all cases the existence
of several solutions for a given value of $\kappa$ has been detected. 

A further generalization is achieved  \cite{BHR05def,BHR05} 
by returning back to D=5 an performing 
a Lorentz boost along the $x^4$ direction. This operation has a non-trivial effect
on the 4D configurations: it produces  
stationary spinning solutions.

\subsection{Non-Abelian braneworlds \cite{Shap03,BBez,BCH,BH04,BBB03}}

Solutions with  gravitating gauge fields can also be considered 
in the context of the braneworld models. 
As was discussed above, the non-Abelian monopole describes a one-dimensional object 
(vortex) in D=5. It will therefore describe a two-dimensional object (domain wall) in D=6 
and a three-dimensional object (3-brane) in D=7. In all cases 
the SU(2) Yang-Mills field and the triplet Higgs fields can be chosen to be  
static and spherically symmetric, 
\be         \label{ans1}
A^a_kdx^k=\epsilon_{akj}\frac{x^j}{r^2}\,
(1-w(r))dx^k,~~~~H^a=\frac{x^a}{r}\,h(r)~~~~/r^2=x^k x^k/,
\ee
with $r^2=x^k x^k$. For the 3-brane in D=7 one chooses  
$x^k$ $/k=1,2,3/$ 
to be coordinates orthogonal to the brane, the 7D metric being
\be            \label{ans2}
ds^2=A(r)\eta_{\mu\nu}dy^\mu dy^\nu+B(r)\delta_{ik}dx^i dx^k\,,
\ee
where $y^\mu$ $/\mu=0,1,2,3/$ are coordinates on the brane. 
Integrating the coupled 
EYM-Higgs-$\Lambda$ equations reveals \cite{Shap03}
that there are solutions satisfying the regularity condition at  the brane,
$A(0)=B(0)=w(0)=1$, $h(0)=0$, for which 
$A(r)\to 0$ for $r\to\infty$. Such solutions describe globally regular braneworlds confined  
in the monopole core with gravity localized on it.   
Solitonic braneworlds have also been studied in systems with a 
gravitating global monopole 
\cite{Shap00} and also  for 
the local and global monopoles \cite{BBez} coupled to each other and to gravity 
in D=7. 

More general solutions of the $n$-brane type have been obtained  \cite{BCH,BH04,BBB03} 
by oxidizing the 
$D=5$ YM vortices to $D=4+n$ dimensions. 
The metric is then chosen to be 
\be            \label{ans3}
ds^2=-e^{\xi_{0}(r)}(dy^0)^2+\sum_{k=1}^{n}e^{\xi_{k}(r)}(dy^k)^2+B(r)\delta_{ik}dx^i dx^k
\ee
with the gauge field given by 
\be               \label{ans4}     
A^a_M dx^M=\epsilon_{akj}\frac{x^j}{r^2}\,(1-w(r))dx^k+\frac{x^a}{r}\sum_{k=1}^{n}h_k(r)dy^k\,.
\ee
What is interesting, the analog of the fixed point \eqref{III} can be obtained \cite{BCH}
within this approach 
 for any $n$, also expressed  in terms of roots of a cubic polynomial.

\section{Non-Abelian solitons in string theory} 

Coming to the question of why one should  study gravitating Yang-Mills fields in 
higher dimensions, one can say that (apart form pure curiosity) the motivation 
for this is provided by string theory. Gravitating Yang-Mills fields enter 
supersymmetry multiplets of the supergravity (SUGRA) theories to which string theory 
reduces at low energies. The knowledge of the basic solutions 
for gravitating gauge fields can therefore be useful for constructing 
solutions in low energy string theory. However, unlike solutions of the pure vacuum gravity,
solutions of the EYM theory will not directly solve equations of SUGRA, since the
latter generically contain additional fields, as for example the dilaton field.  
Constructing supergravity solutions with Yang-Mills fields is 
thus more complicated. However, imposing 
the supersymmetry conditions one can sometimes reduce the problem to solving first order 
Bogomol'nyi equations and not the second order field equations.   

\subsection{Heterotic solitons \cite{Strom90,Duff}} 
The first example of supersymmetric solutions with gravitating Yang-Mills fields 
was obtained by Strominger \cite{Strom90} in the heterotic string theory. The low energy
limit of the latter is a supergravity whose bosonic sector contains a Yang-Mills field
already in D=10. Strominger considers a 5-brane in D=10 and makes the 6+4
split of the metric, 
$$
ds^2=A(x^\mu)\eta_{MN}dy^M dy^N+B(x^\nu)\delta_{\mu\nu}dx^\mu dx^\nu\,,
$$
where $y^M$ /$M,M=0,1,2,3,4,5$/ are coordinates on the brane while the coordinated of the 
4-space orthogonal to the brane are $x^\mu$ $/\mu=1,2,3,4$/. He puts 
the Yang-Mills field to the orthogonal Euclidean 4-space, 
$$
A^a_M=0,~~~~~A^a_\mu(x^\nu). 
$$
Since the Yang-Mills field in D=4 is conformally invariant and the relevant 4D part 
of the metric is conformally flat, the scale factor $B(x^\nu)$ drops out from the 
Yang-Mills equations. As a result, any  flat space self-dual Yang-Mills instanton
in D=4, 
\be                           \label{self}
F_{\mu\nu}=\tilde{F}_{\mu\nu}, 
\ee
will solve the Yang-Mills part of the SUGRA equations. The metric functions $A,B$ as well as
the axion and dilaton will then satisfy the Poisson equation on $\mathbb{E}^4$ with the source 
determined by $F_{\mu\nu}$, and so the solution can be expressed in quadratures. 
As there are many solutions of Eqs.\eqref{self}, this gives a large family
of heterotic 5-branes in D=10. 
Performing then dimensional reductions, one can construct \cite{Duff} yet many more  
different solutions living in  $D<10$. 
All these are called heterotic solitons.

\subsection{Non-Abelian vacua in gauged SUGRAs \cite{CV04,CV97,CV98,CV01,Radu,Radu1}}

Type I and type II string theories and also M theory reduce at low energies to 
supergravities in D=10 and D=11 whose multiplets do not contain Yang-Mills fields.  
However, the latter appear when one performs dimensional reductions on internal
manifolds with non-Abelian isometries. This gives gauged SUGRAs in lower
dimensions whose gauge group is related to the isometry group of the 
internal manifold. For example, the reduction of type II string theory on $S^5$ gives 
the gauged SUGRA in D=5 whose solutions are used in the AdS/CFT correspondence. 
Another example is the reduction of M theory on $S^7$, which 
gives \cite{deWit} the N=8 gauged SUGRA in D=4 with the local 
SO(8)$\times$SU(8) invariance. One can then study solutions 
in these SUGRAs, and lifting them back to $D=10,11$ will give vacua
of string or M theory.  

The field content of a gauged SUGRA model can be quite complicated, which is why 
the gauge fields are often set to zero to find solutions. This gives,
for example, black holes with scalar hair \cite{Duff-Liu,Hertog-Maeda} in N=8 SUGRA. 
However, in some cases one can construct solutions also 
with gravitating gauge fields.  Let us consider a SUGRA model \cite{CV04} in D=4 which is a 
consistent truncation of the maximal N=8 SUGRA,
\be               \label{so4}
{\cal L}_4  =\frac{1}{4}R
-\frac12\partial_\mu\phi\partial^\mu\phi
-\frac{1}{4}\,e^{2\phi}F^{a}_{\mu\nu}F^{a\mu\nu}
+\frac{1}{8}\, (e^{-2\phi}+{\xi^2}e^{2\phi}+4{\xi}).
\ee
This theory contains a gravity-coupled Yang-Mills field with gauge group SU(2)
and the dilaton, whose potential depends on a real parameter $\xi$. 
In the static and spherically symmetric case the fields are given by 
\bea 
ds^2_{(4)}&=&-e^{2V(\rho)}dt^2+e^{2\lambda(\rho)}d\rho^2+r^2(\rho)d\Omega^2, \nonumber \\
A^a_k dx^k&=&(w(\rho)-1)\epsilon_{aik}n^idn^k,~~ ~\phi=\phi(\rho).~~\nonumber 
\eea 
For one-shell configurations the amplitudes $V,\lambda,r,w,\phi$ satisfy 
a system of second order ODEs whose solutions can be studied \cite{Mann} numerically. 
Instead of solving these equations, one can also consider the 
conditions for the fields to have unbroken supersymmetries.  
These conditions require the existence of a non-trivial spinor 
$\epsilon$ satisfying the linear equations 
\bea                   \label{SUSY}
0&=&
\frac{1}{\sqrt{2}}\,\gamma^\mu \partial_\mu\phi\,\epsilon+
\frac{1}{2}\,e^{\phi}{\cal F}\epsilon
+\frac{1}{4}\,(e^{-\phi}-{\xi} e^{\phi})\epsilon, \nonumber \\
0&=&
{\cal D}_\mu\epsilon +
\frac{1}{2\sqrt{2}}\,e^{\phi}{\cal F}\gamma_\mu \epsilon 
+\frac{1}{4\sqrt{2}}\,(e^{-\phi}+{\xi} e^{\phi})\gamma_\mu\epsilon,  \label{susy}
\eea
where ${\cal F}=\frac{i}{2}\tau^aF^a_{\alpha\beta}\gamma^\alpha\gamma^\beta$. 
In general these equations are inconsistent, since there are 80 equations for 
16 components of $\epsilon$, and so the only solution is $\epsilon=0$. 
However, one can show \cite{CV04} that if the background fields are such that the following
conditions are fulfilled, 
\bea             \label{bog}
&&V'-\phi'={ \xi}\,\frac{P}{\sqrt{2}N}e^{\phi+\lambda},~~~~~
Q=e^{V+\phi}\frac{w}{N}, ~~~~~ \phi'=\sqrt{2}\,\frac{BP}{N}\, e^{\lambda},\nonumber \\
w'&=&-\frac{rwB}{N} e^{-\phi+\lambda},~~~~~
N\equiv\rho\lambda'+1=\sqrt{w^2+P^2},~~~~~~
r'=Ne^\lambda,  
\eea
where $Q$ is a constant and 
$$
P=e^\phi\frac{1-w^2}{\sqrt{2}r}+\frac{r}{2\sqrt{2}}(e^{-\phi}+{\xi} e^\phi),~~~~
B=-\frac{P}{\sqrt{2}r}+\frac12e^{-\phi},
$$   
then there exist four independent solutions of Eqs.\eqref{susy}, 
which corresponds to the N=1 supersymmetry. 
\begin{figure}[h]
\hbox to\linewidth{\hss%
  \resizebox{6.5cm}{5.5cm}{\includegraphics{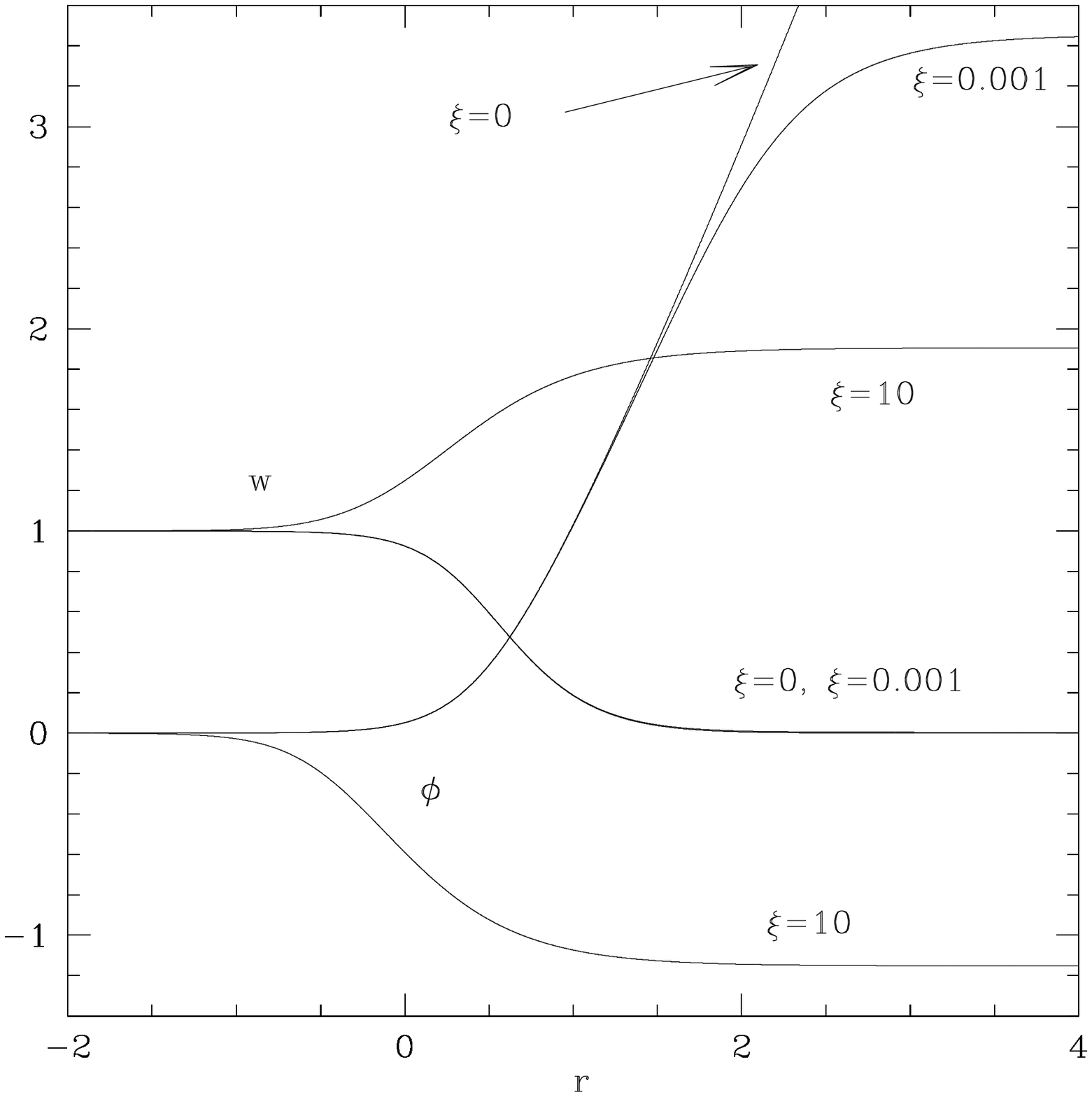}}%
  \resizebox{6.5cm}{5.5cm}{\includegraphics{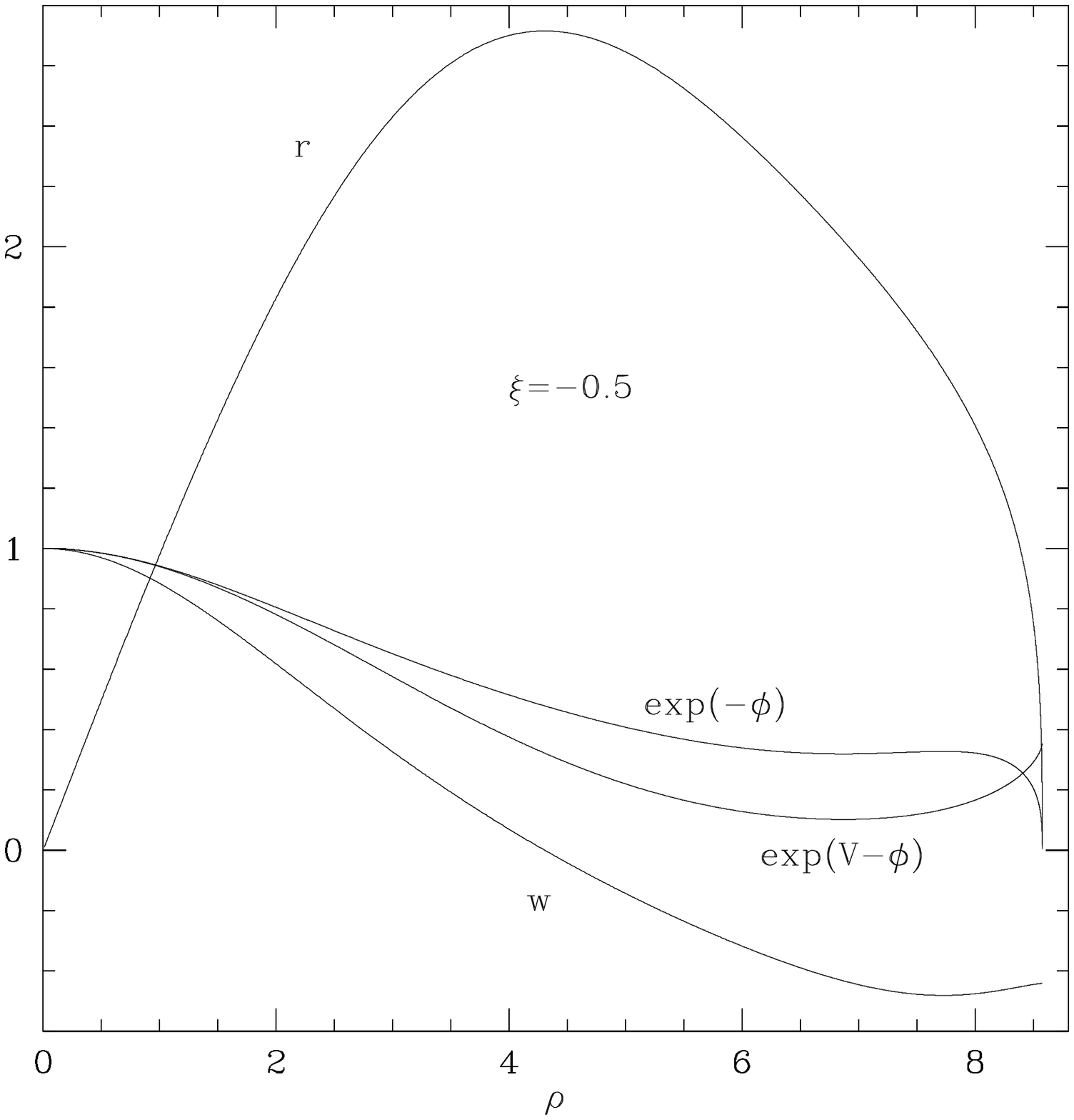}}%
\hss}
\caption{The asymptotically AdS and compact solutions of the Bogomol'nyi equations \eqref{bog}.
}
\end{figure}
Solutions \cite{CV04} 
of the Bogomol'nyi equations \eqref{bog} with the regular boundary condition
at the origin, $\rho=r=0$, comprise a continuous family labeled by $\xi$ and 
show three completely different types of behavior,
depending on the sign on $\xi$. For $\xi>0$ the dilaton is everywhere bounded
and the solutions approach the AdS metric in the asymptotic region. 
For $\xi<0$ the solutions are of the `bag of gold' type, since they have compact spatial 
sections with the topology of $S^3$. The geometry is generically singular at one pole
of the $S^3$. However, for $\xi=-2$ the solution is globally regular, the dilaton is 
constant, the geometry is that of $\mathbb{R}^1\times {S}^3$, and the gauge field potential
coincides with the invariant forms on $S^3$. All these solutions can be uplifted \cite{CV04} 
to $D=11$ to become vacua of M theory.

For $\xi=0$ the solution is given by \cite{CV97} 
\bea                                    \label{4.7}
d{ s}^{2}&=&\left.\left.
2\, e^{2\phi}\,
\right\{
-dt^{2}+d\rho^{2}+
\left.\left.\R^{2}(\rho)\right(d\vartheta ^{2}+\sin ^{2}\vartheta d\varphi
^{2}\right)\right\} , \nonumber                 \\
w&=&\pm \frac{\rho}{\sinh \rho},\ \ \ \
e^{2(\phi-\phi_0) }=\frac{\sinh \rho}{2\,R(\rho)},~~~
\R(\rho)=\sqrt{2\rho\coth\rho-w^2-1}.            \label{sol}
\eea
This solution can be uplifted \cite{CV98}  to D=10, the string frame metric
and the three-form in D=10 then being given by 
\bea
d{s}^2 &=&
-{ dt^{2}+dy_1^2+dy_2^2+dy_3^2}+d\rho^{2}
+\R^{2}(\rho)\, d\Omega^{2}_2
+\Theta^a\Theta^a\, , \nonumber \\
H&=&                        \label{sol1}
\frac{1}{2\sqrt{2}}\, e^{-\frac{3}{4}\phi}(F^{a}\wedge\Theta^a
+
\epsilon_{abc}\Theta^a\wedge\Theta^b\wedge\Theta^c),
\eea
where $F^a$ is the gauge field 2-form and 
$\Theta^a=A^a-\theta^a$ with $\theta^a$ being the invariant forms on $S^3$. 
This solution describes a 3-brane in D=10.

This solution was originally obtained \cite{CV97,CV98} by
Chamseddine and Volkov, both in D=4 and D=10. A very interesting holographic 
interpretation 
for this solution was then proposed \cite{MN} by Maldacena and Nunez,   
according to which this solution describes the NS-NS 5-brane wrapped on $S^2$,
in which case it effectively becomes a 3-brane. 
As a result, the solution provides a dual SUGRA description for the 
wrapped brane worldvolume theory, which is the (deformed) N=1 Super Yang-Mills
in D=4. 
Since this 
theory is confining, one can say that the solution \eqref{sol},\eqref{sol1}
provides the dual SUGRA description for the phenomenon of confinement. 
For example, the value of the Wilson loop, the beta-function and other parameters 
of the confining 
gauge field theory can be obtained by simply computing some purely geometrical 
parameters of the solution, like areas of spheres and the values of the $H$-fluxes 
through them. As a result, this solution with self-gravitating 
Yang-Mills field has found quite interesting and serious applications. 
Generally known as solution of Maldacena and Nunez 
(the names of  Chamseddine and Volkov seem now to be completely forgotten) 
it plays an important role in the analysis of string theory 
(see \cite{Grana,Edelstein} for recent reviews).

Non Abelian vacua have also been studied \cite{CV01} in the context of a gauged SUGRA 
in D=5. 
The gravitational and gauge fields in this case are given by the same SO(4) invariant 
expressions \eqref{e0} as for the YM particle, and there are also the dilaton and axion. 
The solutions are not asymptotically flat, they have two supercharges, and they can be 
used \cite{MN1} 
for a dual SUGRA description of the confining N=1 Super-Yang-Mills theory in D=3.  

The discussed above supersymmetric solutions \cite{CV97,CV01} have also been generalized 
\cite{Radu,Radu1} to the case
where the spatial part of the metric is $\mathbb{R}^1\times \Sigma$, where $\Sigma$ is a maximal
symmetry space, that is sphere, hyperboloid or Euclidean space. 
Their black hole generalizations \cite{GTV,Bertoldi} have been used 
for a holographic description of the confinement/deconfinement phase transition 
in the dual gauge theory.

\end{document}